\documentclass{jpsj3}
\usepackage{txfonts}
\usepackage{comment}
\usepackage{bm}

\title{
Fluctuation of a bilayer composed by amphipathic molecules
}

\author{Shunta Kikuchi and Hiroshi Watanabe\thanks{hwatanabe@appi.keio.ac.jp}}
\inst{Faculty of Science and Technology, Department of Applied Physics and Physico-Informatics, Keio University, Yokohama 223-8522, Japan} 

\abst{
    We investigated the properties of bilayer fluctuations using molecular dynamics. We modeled the amphipathic molecules as a diatomic molecule, constructed a bilayer in the solvent, and observed the Fourier spectrum of its fluctuations. The results showed that $q^4$ behavior was dominant at the high temperature, whereas a crossover from $q^4$ to $q^2$ was observed at the low temperature. This behavior is complementary to the results for monolayer membranes, where $q^2$ behavior was dominant when the interfacial tension was high, whereas a crossover from $q^2$ to $q^4$ was observed when the interfacial tension was lowered~[S. Kikuchi and H. Watanabe, J. Chem. Phys., $\bm{158}$ 12 (2023)]. These results suggest that the restoring force is dominated by elasticity at high temperatures and by interfacial tension at low temperatures. We also observed radial distribution functions to investigate the structure of amphipathic molecules in bilayers. The structure of the amphipathic molecules depends on temperature in the direction horizontal to the interface, while it does not depend on temperature in the direction perpendicular to the interface. This result suggests that bilayers increase fluctuations while maintaining the membrane structure in high temperature.
}

\begin{document}
\maketitle

\section{Introduction}\label{sec:intro}

The lipid bilayer consists of two layers of lipids, with the hydrophilic heads facing outward and the hydrophobic tails facing inward\cite{nagle2000structure,bloom1991physical}. Since the lipid bilayer is a fundamental structure forming the basis of the cell membrane, understanding lipid bilayers is essential both in biology and drug delivery\cite{peetla2009biophysical}. However, describing lipid bilayers using the continuum representation is challenging since the atoms' behavior significantly affects the membrane's properties. Recently, developments of computational power allow us to perform all-atom simulations of membranes by molecular dynamics (MD) simulations\cite{goetz1998computer,goetz1999mobility}. The numerical studies investigated the generation process of the vesicles\cite{marrink2003molecular, Shinoda2010}, the relation between the fluidity and the molecular orientation\cite{coppock2009atomistic,venable1993molecular}, and the drug permeation through lipid bilayers.\cite{orsi2010permeability}.

One of the fundamental properties of the membrane is its elasticity. For example, when the cell is deformed, the elasticity of the bilayer maintains its shape\cite{fabrikant2009computational}. In the ventricular assist devices, the local pressure gradient causes hemolysis with damage in the red blood cells\cite{viallat2014red,yu2017review}. In addition to the deformation, the lipid bilayers are affected by the blood flow and the thermal fluctuations\cite{de1997deformation,zgorski2019surface,enkavi2019multiscale,leonard2019developing}. Helfrich studied the theory of membrane elasticity, revealing that the fluctuation spectra of the membranes are described by the interfacial tension and membrane elasticity\cite{helfrich1973elastic}. The properties of the membranes have been investigated by comparing experimental and simulation results with Helfrich's theory.

Coarse-grained lipid bilayer models are often adopted in molecular dynamics simulations to reduce computational complexity\cite{noguchi2009membrane,venturoli2006mesoscopic,marrink2019computational}. Implicit solvents are also used because of separation in the time scales of motion between lipid bilayers and water atoms\cite{cooke2005tunable,brannigan2005flexible}. In these studies, the parameters are set to reproduce the physical properties of real lipid bilayers, and only the elastic $q^4$ behavior is observed in the Fourier spectrum of the membrane\cite{brannigan2005flexible, Brandt2011}. On the other hand, the real lipid molecules are highly complex, causing the entanglement of hydrophobic groups and the formation of interdigitated structures. Therefore, it is difficult to clarify which structures and properties of lipid molecules are attributable to the physical phenomena, even in the case of coarse-graining.

In the previous study, we investigated the properties of the monolayer spontaneously formed by diatomic amphiphilic molecules to clarify the relationship between molecular structure and membrane mechanical properties\cite{Kikuchi2023}. In the monolayer case, the membrane's main restoring force was found to originate from the interfacial tension with finite interfacial tension. In contrast, the crossover from the interfacial tension to the elasticity was observed as the interfacial tension decreased. To clarify the relationship between the bilayer and the interfacial tension, we investigated the mechanical properties and structure of the bilayer spontaneously formed by diatomic molecules, which is the simplest model for representing amphiphilic molecules. In addition, there can be differences in behavior derived from the structure of monolayers and bilayers, but how they differ needs to be clarified. In this paper, we discuss the similarities and differences in the behavior of monolayers and bilayers by molecular dynamics simulations.

The rest of the paper is organized as follows. In section \ref{sec:method}, we describe the method. The results are described in Sec.~\ref{sec:results}. Section~\ref{sec:summary} is devoted to the summary and discussion.

\section{Method}\label{sec:method}

We adopted the Lennard-Jones (LJ) and the Weeks-Chandler-Andersen (WCA) potential as interactions of amphipathic molecules. The LJ potential is described by
\begin{equation}
	\phi(r)=\left\{
	\begin{array}{ll}
		4 \varepsilon \left \{ \left  ( \frac{\sigma}{r} \right )^{12} - \left ( \frac{\sigma}{r} \right )^{6} \right \} & (r < r_c),     \\
		0                                                                                                                & (r \geq r_c ),
	\end{array}
	\right.
	\label{eq:lj}
\end{equation}
where $\sigma$ is the atomic diameter, $\varepsilon$ is the well depth and $r$ is the distance between the atoms. The cutoff length is denoted by $r_c$. In the following, we adopt the reduced unit such that the $\sigma, \varepsilon$, and $k_\mathrm{B}$ are unity. We set $r_c = 3.0$ for the LJ potential and $r_c = 2^{1/6}$ for the WCA potential. The LJ interaction represents the attraction between atoms while the WCA represents the repulsion.

\begin{figure}[htbp]
	\centering
	\includegraphics[scale = 0.22]{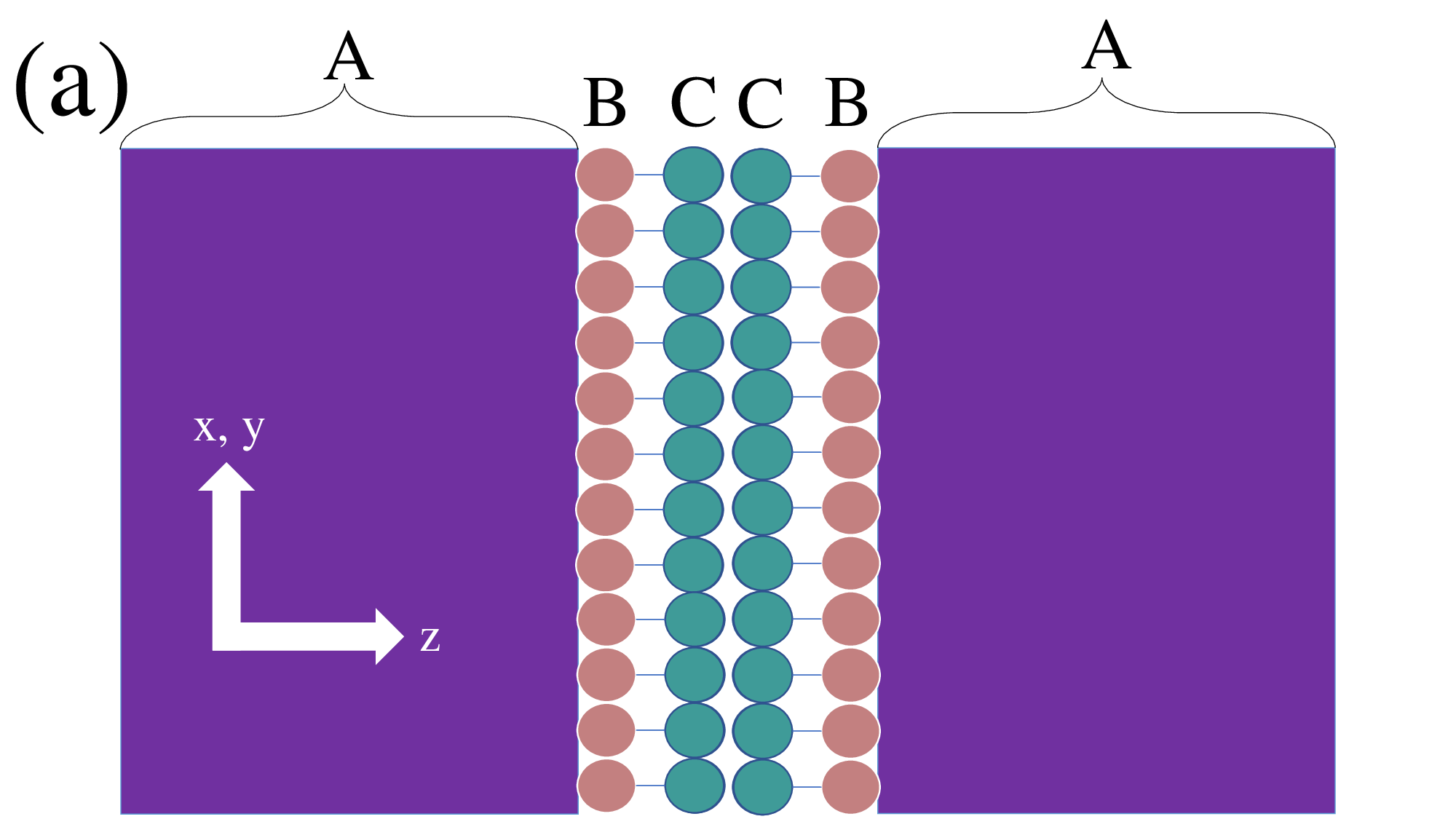}
	\includegraphics[scale = 0.22]{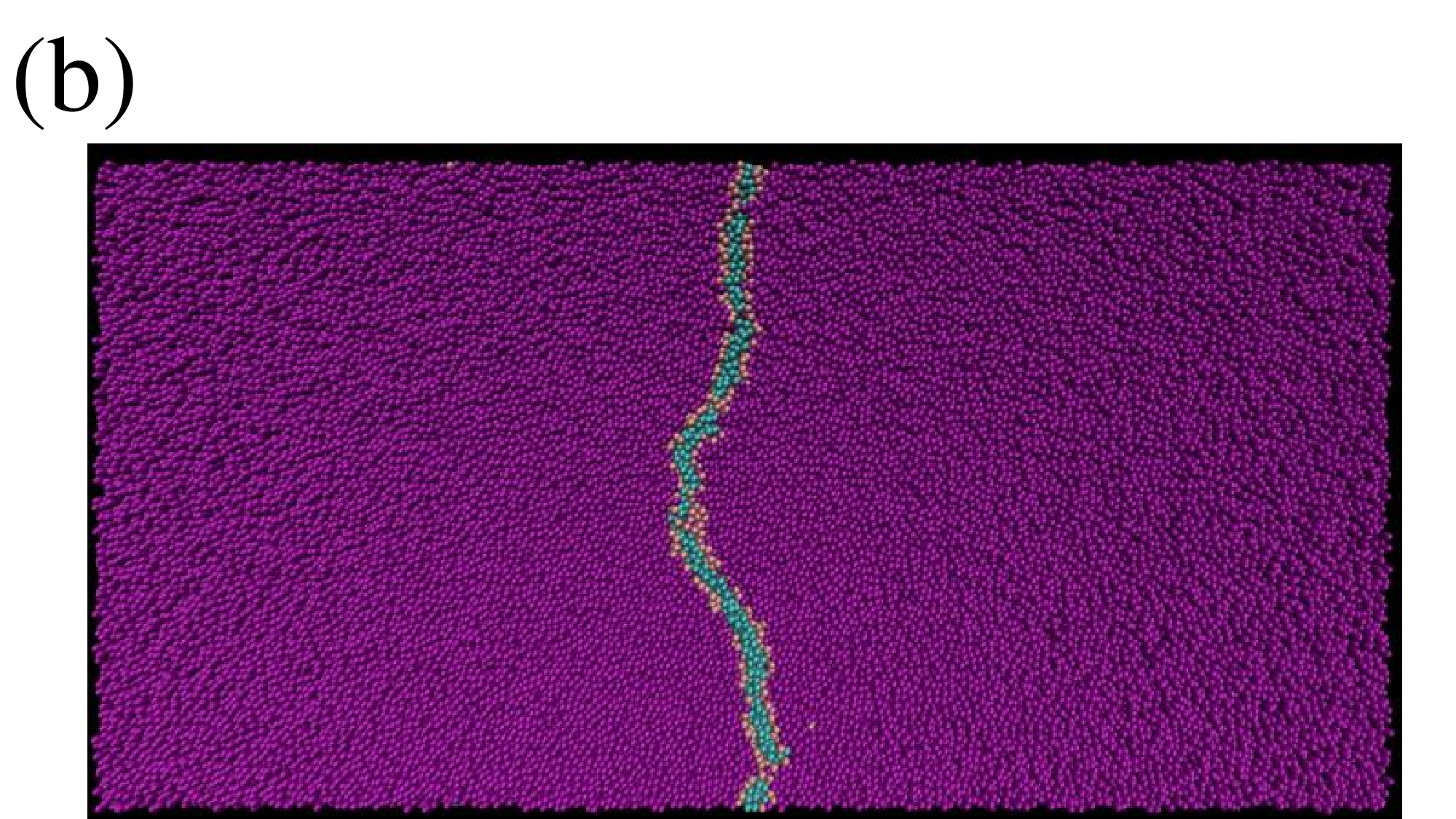}
	\caption{(Color online) (a) The schematic view of the simulation box. The bilayer membrane separates the solvent. The interface is normal to the $z$-axis. The solvent, hydrophilic, and hydrophobic atoms are referred to as A, B, and C atoms, respectively. (b) The typical snapshot of the simulations. The interface fluctuates.}
	\label{fig:model_bilayer}
\end{figure}

In this paper, the solvent molecules, the hydrophilic and hydrophobic molecules of the amphipathic molecules were represented as one atom, respectively. The amphipathic molecule is modeled as a diatomic molecule with a hydrophobic atom and a hydrophilic atom bonded together. In order to investigate the fluctuations of the membrane, we prepare the bilayer separating the solvent as shown in Fig.~\ref{fig:model_bilayer}~(a). The amphipathic molecule are placed at the liquid-liquid interface and act as the amphiphilic molecules. The $z$ axis is the normal to the interface. As shown in Fig.~\ref{fig:model_bilayer}~(b), the solvent, the hydrophilic, and the hydrophobic atoms are referred to as A, B, and C atoms, respectively. We set the interaction between the solvent atoms and the hydrophilic atoms (A-B) as the LJ potential, the interaction between A-C as the WCA potential, and the interaction between B-C as the WCA potential with the well depth $\varepsilon = 1.05$. The interaction between the same atoms is set as the LJ potential. We adopted the harmonic potential as the bond interaction in the amphipathic molecule which is
\begin{equation}
	V(r) = K(r-l_0)^2
	\label{eq:harmonic}
\end{equation}
where $l_0$ is the bond length and $K$ is the spring constant. The bond length is set to $l_0 = 1.50$ and the spring constant is set to $K = 200$ throughout this study. Note that the interactions between atoms of the same amphipathic molecules are not considered.

The typical snapshot is shown in Fig.~\ref{fig:model_bilayer} (b). We used Visual Molecular Dynamics (VMD) for the visualization \cite{humphrey1996vmd}. The total number of the atoms was set to $N = 1~600~000$ including $16~000$ amphipathic molecules. Since each amphipathic molecule contains two atoms, there are $8~000$ molecules in the system. The simulations were performed using the Large-scale Atomic/Molecular Massively Parallel Simulator (LAMMPS) \cite{LAMMPS1995}. The temperature and pressure in the simulation box are maintained by the Nos\'e-Hoover thermostat and the Andersen barostat\cite{nose1984molecular,hoover1985canonical,andersen1980molecular}. The pressure was set to $P = 0.20$ and the temperature range $0.65 \leq T \leq 0.80$ was studied. The velocity Verlet algorithm is used for the time integration with a time step of $0.005$\cite{verlet1967computer}. Simulations were performed up to $3~000~000$ steps.

After the system had equilibrated, we calculated the spectra of membrane fluctuations and the radial distribution function (RDF) $g(r)$ of atoms. The area near the interface ($80<z<120$) of the simulation box was divided into $G \times G$ sections in the $x-y$ direction. The center of gravity of the amphipathic molecule in each rectangle was defined as the interface position $h(x,y)$, where $G=30$ in this study. Then we obtain the spectrum $h(q)$ by the Fourier transformation of $h(x,y)$, where $q$ is the wavenumber. The Fourier transform of the height $h(x,y)$ is given by
\begin{equation}
	h(q_x,q_y) = \frac{1}{G} \sum_{x=0}^{G-1}\sum_{y=0}^{G-1} h(x,y) \exp \{ -i2\pi(q_xx+q_yy) /G \}.
\end{equation}
After calculating the $h(q_x,q_y)$, we obtain the $h(q)$ by averaging $h(q_x,q_y)$ for the angular distribution, where $q=\sqrt{q_x^2+q_y^2}$. The spectra of fluctuations are denoted by\cite{goetz1998computer}
\begin{equation}
	|h(q)|^2 = \frac{k_\mathrm{B} T}{\gamma q^2+\kappa q^4}.
	\label{eq:fluctuation}
\end{equation}
where $\gamma$ is the interface tension, $\kappa$ is the elasticity of the interface, $k_\mathrm{B} = 1$ is the Boltzmann constant, respectively. Equation~(\ref{eq:fluctuation}) means that the $q^2$ component of the membrane fluctuation originates from the interfacial tension and the $q^4$ components from the elasticity.

The usual definition of the RDF is normalized by the average density so that it becomes $1$ in the far distance. However, this definition results in unphysically large values in short distance since the amphiphilic molecules are localized in the membrane. Therefore, we adopt the unnormalized RDF, which is
$$
	g_{XY}(r) = \frac{2n_{XY}(r)V}{N_X+N_Y},
$$
where $ n_{XY} (r) $ is the number of Y-atoms within distance $ r$ to $ r+ dr$ from X-atoms, $ N_X$ and $ N_Y$ are the total numbers of X- and Y-atoms, $ dr$ is the size of bins, and $ V$ is the volume of the simulation box, respectively. Since the value of the unnormalized RDF is the number density in each bin, the definition allows us to compare RDFs in different atomic species.

\section{Results}\label{sec:results}

\begin{figure}[htbp]
	\centering
	\includegraphics[scale = 0.3]{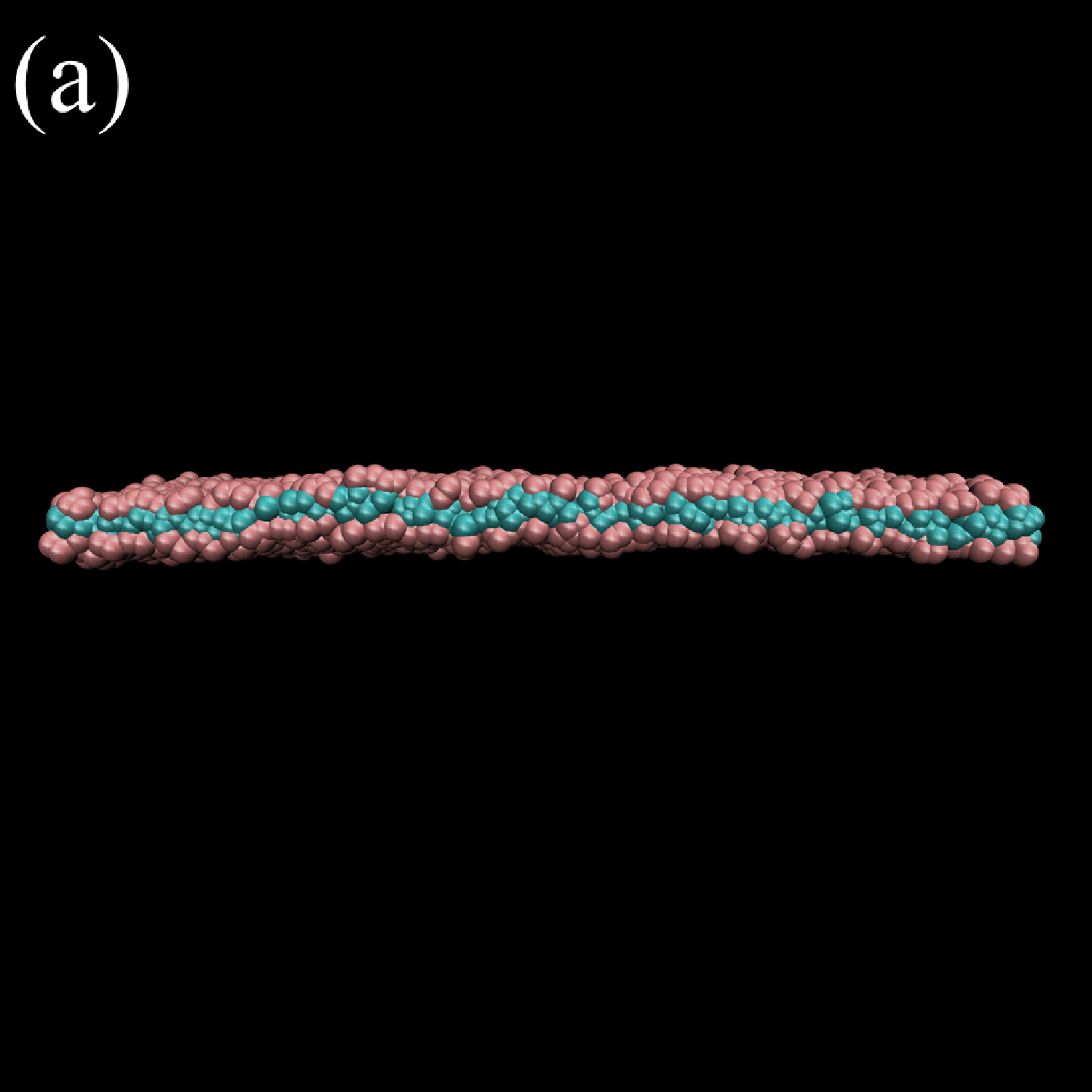}
	\includegraphics[scale = 0.3]{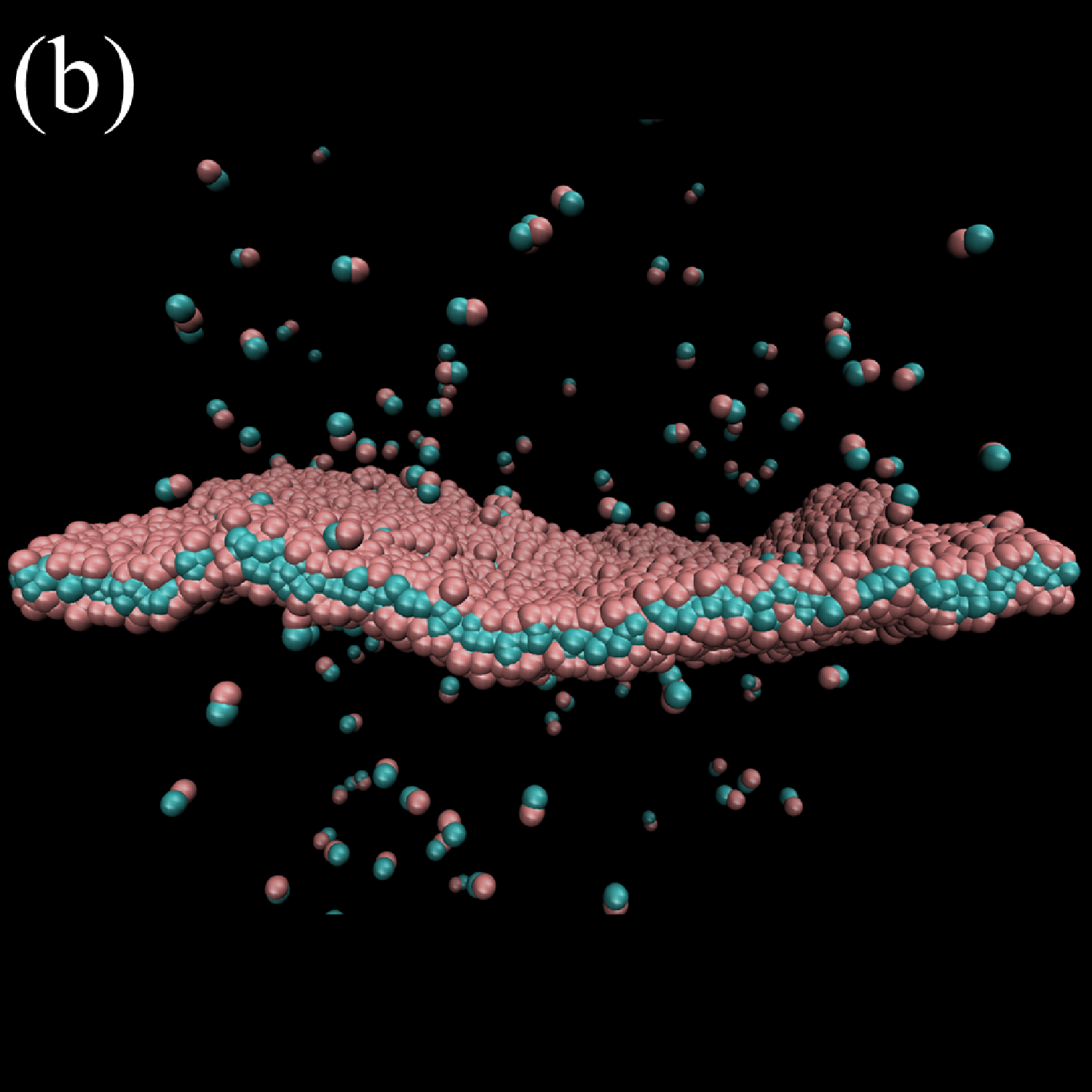}
	\caption{(Color online) The snapshots of the simulations in (a) low temperature ($T=0.70$) and (b) high temperature ($T=0.80$). Only amphipathic molecules are shown for the visibility. While the membrane is almost flat in low temperature, it is fluctuating in high temperature.
	}
	\label{fig:snapshots}
\end{figure}

The snapshots of the simulations are shown in Fig.~\ref{fig:snapshots}. The membrane was flat in low temperature ($T=0.70$), while the membrane fluctuated significantly in high temperature ($T = 0.80$). Therefore, it is expected that the fluctuation properties of the interface depend on temperature.

Note that the amphiphilic molecules are significantly dissolved in the solvent at high temperatures, as shown in Fig.~\ref{fig:snapshots}~(b). This phenomenon is due to the fact that we adopted diatomic molecules as amphipathic molecules, which will not happen for actual lipids since they are highly hydrophobic due to their long hydrophobic tails. However, the effect of the dissolution into the solvent does not affect the membrane's mechanical properties once equilibrium is reached.

\begin{figure}[htbp]
	\centering
	\includegraphics[scale = 1]{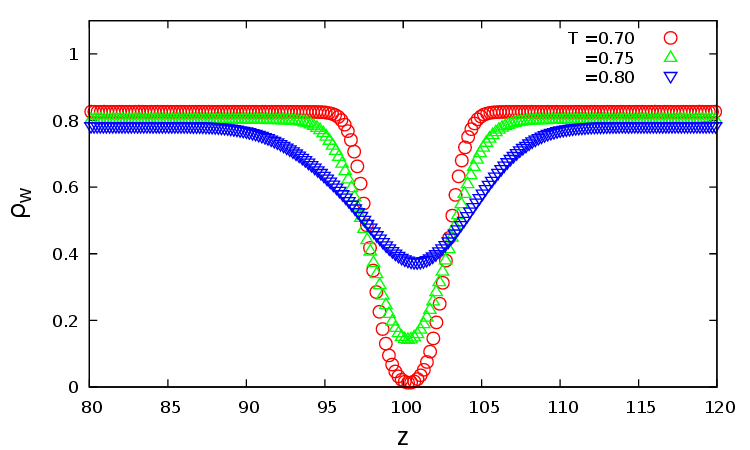}
	\caption{The solvents' density profile. The density is almost zero at low temperature ($T = 0.70$). Therefore, the membrane completely separates the interface. The density of the solvents at the interface increases as the temperature increases.
	}
	\label{fig:density}
\end{figure}

The density profile of the solvent atoms is shown in Fig.~\ref{fig:density}. At the high temperatures, solvent densities were finite near the interface, and solvent appeared to seep into the membrane. However, this behavior originated from membrane fluctuations and not the seeping of solvent molecules, as will be shown later.

\begin{figure}[htbp]
	\centering
	\includegraphics[scale=1]{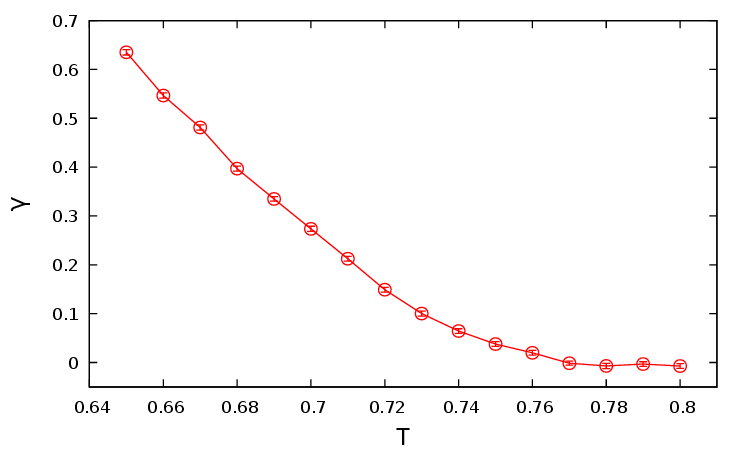}
	\caption{The temperature dependence of the interfacial tension. The interfacial tension decreases as the temperature increases and becomes virtually zero at $T=0.80$.}
	\label{fig:itf}
\end{figure}

In the previous work on the fluctuations of the monolayer membrane, the restoring force originated from the interfacial tension when the interfacial tension was finite, whereas the elasticity became the dominant force when the interfacial tension was virtually zero\cite{Kikuchi2023}. This investigation suggests that the fluctuation's nature also changes with the membranes' interfacial tension. To study the relation between the surface tension and the fluctuation of the membrane, we investigated the temperature dependence of the interface tension, which is shown in Fig.~\ref{fig:itf}. At the low temperature ($T=0.70$), when the membrane was almost flat, the interfacial tension was finite, while at the high temperature ($T=0.80$), when the membrane was highly fluctuating, the interfacial tension was virtually zero.

\begin{figure}[htbp]
	\centering
	\includegraphics[scale = 0.6]{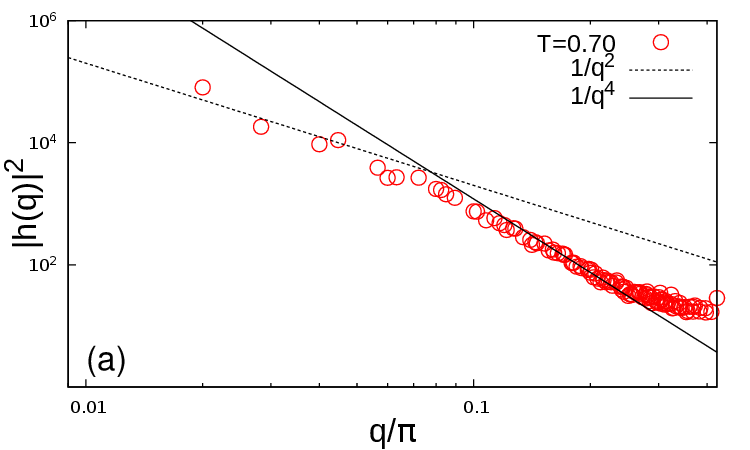}
	\includegraphics[scale = 0.6]{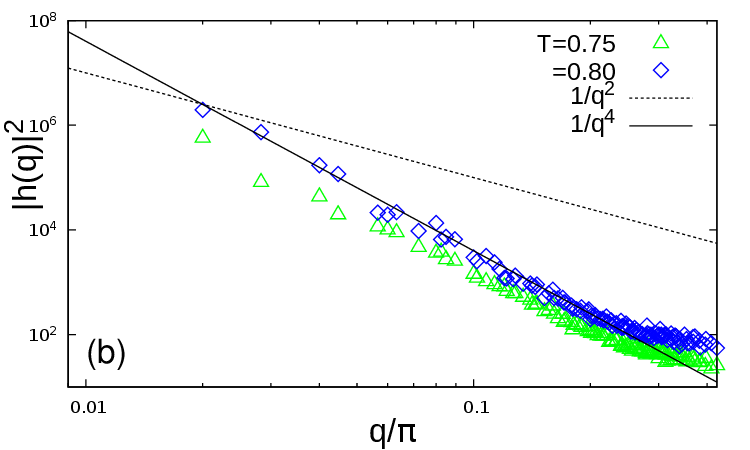}
	\caption{The temperature dependence of the spectra $|h(q)|^2$. The Decimal logarithm is taken for both axes. (a) The low temperature ($T=0.70$). $q^4$ was only observed. (b) The high temperature ($T=0.75, 0.80$). The crossover from $q^2$ to $q^4$ was observed.
	}
	\label{fig:fluc}
\end{figure}

\begin{figure}[htbp]
	\centering
	\includegraphics[scale=0.6]{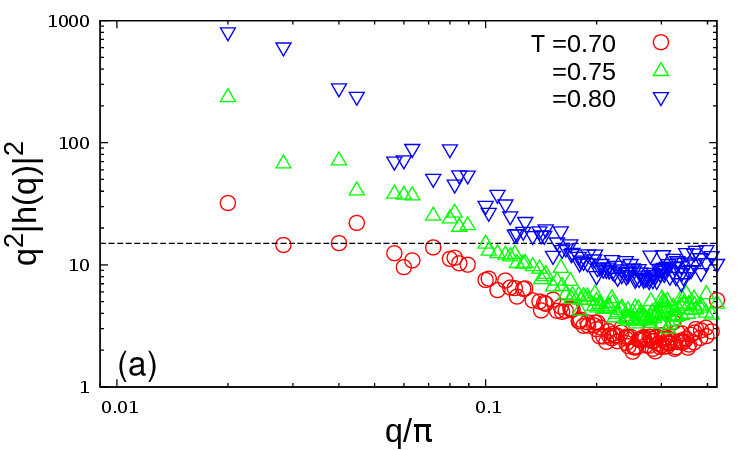}
	\includegraphics[scale=0.6]{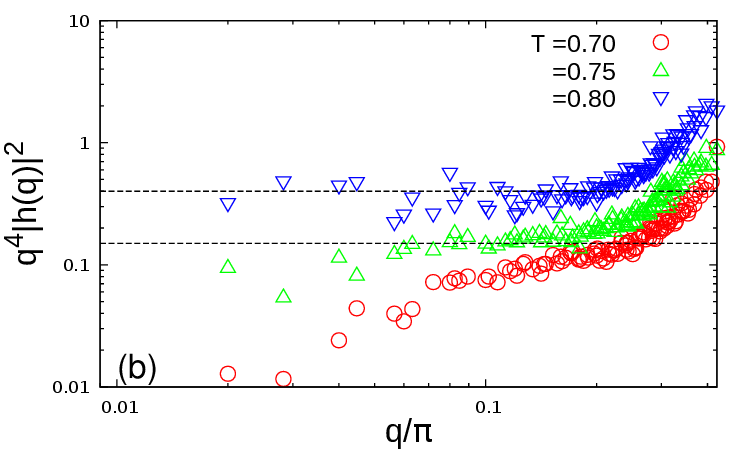}
	\caption{Graphs of (a) $q^2|h(q)|^2$ and (b) $q^4|h(q)|^2$. The Decimal logarithm is taken for both axes. At the low temperature ($T = 0.70$), $q^2|h(q)|^2$ was constant in the small wavenumber region. This behavior indicates that $q^2$ is dominant. With the high temperature ($T = 0.80$), $q^4|h(q)|^2$ was constant in the small wavenumber region. This behavior indicates that $q^4$ is dominant.}
	\label{fig:q_fluctuation}
\end{figure}

\begin{figure}[htbp]
	\centering
	\includegraphics[scale=0.6]{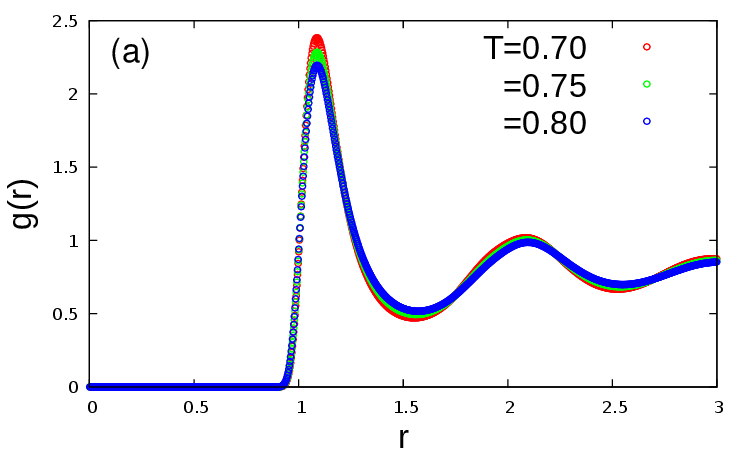}
	\includegraphics[scale=0.6]{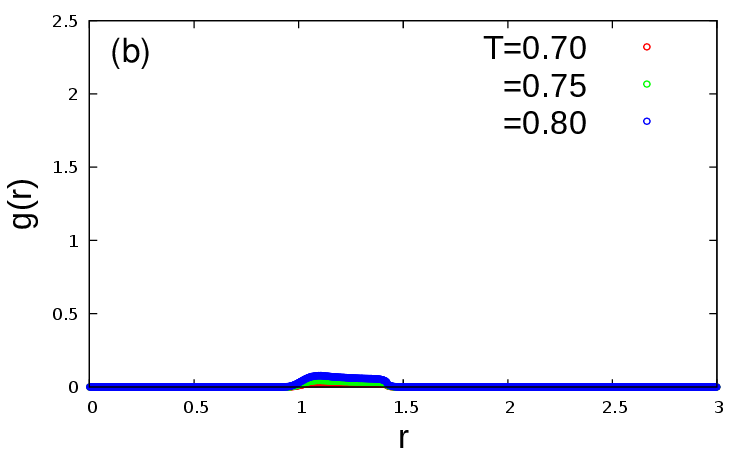}
	\caption{Graphs of the RDF $g(r)$. (a) The RDF between the solvent atoms (A-A). The temperature dependence was seen. (b) The RDF between the solvent atoms and the hydrophobic atoms (A-C). The RDF was almost zero, which indicates that the solvent did not penetrate the membrane.}
	\label{fig:rdf1}
\end{figure}

\begin{figure}[htbp]
	\centering
	\includegraphics[scale=0.6]{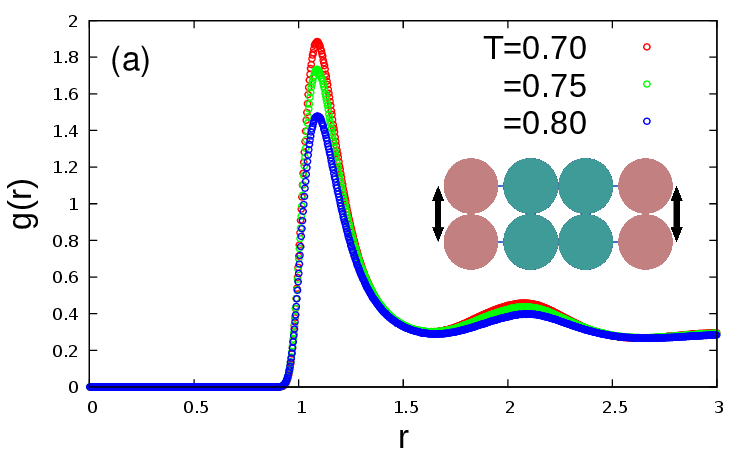}
	\includegraphics[scale=0.6]{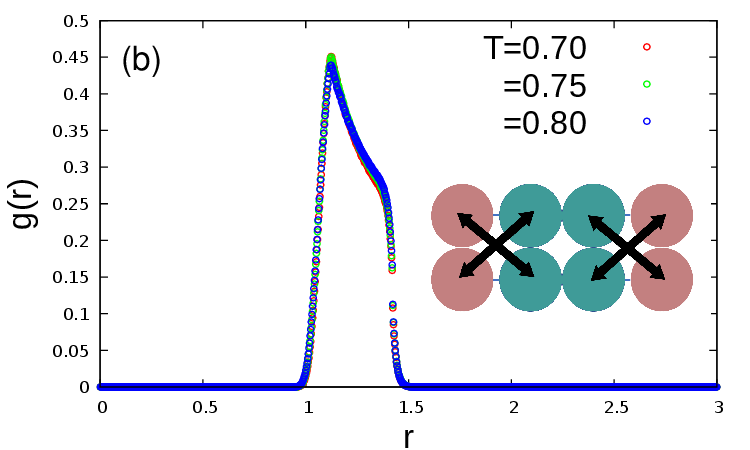}
	\caption{Graphs of the RDF $g(r)$. (a)The RDF between the hydrophilic atoms (B-B). There exists the temperature dependence. (b) The RDF between the hydrophilic atoms and the hydrophobic atoms (B-C). The RDFs exhibit virtually no temperature dependence.}
	\label{fig:rdf2}
\end{figure}

In the case of the monolayer membrane, the crossover from $q^2$ to $q^4$ was observed as the interface tension decreased, which reflects that the restoring force was changed from the interfacial tension to the elasticity. Therefore, we investigated the fluctuation spectra of the membrane to identify its restoring force. The temperature dependence of the fluctuation spectra is shown in Fig.~\ref{fig:fluc}. While the crossover from $q^2$ to $q^4$ was observed at the low temperature, only $q^4$ behaviors were observed at high temperatures.
To identify the power exponents of the spectra, $q^2|h(q)|$ and $q^4|h(q)|$ are plotted as functions of $q$ in Fig.~\ref{fig:q_fluctuation}\cite{watson2012determining}. The $q^2$ behavior exists at low temperatures, but the $q^4$ behavior becomes dominant at high temperatures.
These results indicate that $q^2$ behavior was observed when the interfacial tension was finite, and only $q^4$ behaviors were observed when the interfacial tension was virtually zero. The change in the fluctuation can be originated from the change in the crystalline structure of the membrane. For example, the membrane can be stiff at a low temperature since the amphipathic molecules are aligned in an interdigitated crystal structure. Still, the membrane becomes soft since the crystalline structure melts as temperature increases. To investigate the crystal structure, we calculated the RDF of the membrane. The RDF between solvents (A-A) is shown in Fig.~\ref{fig:rdf1} (a), and the RDF between solvent and hydrophobic atoms (A-C) is shown in Fig.~\ref{fig:rdf1} (b). The temperature dependence was observed in the RDF between A-A. This result is consistent with the general solvent.
On the other hand, the RDF between A-C was almost zero, regardless of the temperature. This behavior indicates that the solvent atoms did not penetrate the membrane, and the interaction between the solvent and the hydrophobic atoms did not change the interfacial tension. Figure \ref{fig:rdf2} (a) shows the RDF between the hydrophobic atoms (B-B), and Fig.~\ref{fig:rdf1} (a) shows the RDF between B-C. The apparent temperature dependence was observed in the RDFs between the same types of atoms. These results imply that the horizontal fluidity in the membrane increases as the temperature increases.
On the other hand, the temperature dependence was not observed in the RDF between the hydrophilic atoms and the hydrophobic atoms. This result shows that the structure of the membrane did not change with temperature. Consequently, the membrane increases its fluctuation, keeping its microscopic structure as the temperature increases.


The fluctuations may change due to structural changes in the membrane, such as the formation of interdigitated structures. However, if the bilayer had an interdigitated structure, a peak would appear at the high wavenumber end of the Fourier spectrum in Fig.~\ref{fig:fluc}, but this peak is not observed. In addition, a characteristic peak due to the interdigitated structure should be visible in Fig.~\ref{fig:rdf2}, but only a liquid-like structure is observed. Therefore, we concluded that the amphiphilic molecule exists as a two-dimensional liquid. However, thermal fluctuations may bury the Fourier peak at the high wavenumber side. Since our model is diatomic, the effect of the interdigitated structure on the RDF may be small. A molecular model with more extended hydrophobic groups would be required to identify the presence of an interdigitated structure.

\section{Summary and Discussion}\label{sec:summary}

In this paper, we investigated the fluctuations of bilayer membranes using MD simulations. The power spectra of the membrane fluctuations changed with temperature. Since the membrane separated the solvents, the interfacial tension was slightly affected, and $q^4$ behavior was only observed with low temperatures. However, when the temperature was low enough to see the effect of the interfacial tension, the crossover from $q^4$ to $q^2$ behaviors was observed.

Here, we discuss the relation between the monolayer and the bilayer membranes. Table~\ref{table:crossover} summarizes the relation between the fluctuations and the interfacial tension, combined with the results of the monolayers\cite{Kikuchi2023}. In the previous study, $q^2$ behavior was observed, and the interfacial tension was dominant for the fluctuations of monolayer when the interfacial tension was finite, whereas the crossover from $q^2$ to $q^4$ was observed, which indicates that the elasticity was dominant when the interfacial tension was almost zero. In this paper, increasing the temperature in the bilayer membrane corresponds to reducing the interfacial tension, which is complementary to the results for the monolayer.

\begin{table}[htbp]
	\begin{center}
		\caption{Table of fluctuations in monolayer and bilayer membranes. In the monolayer membrane, $q^2$ behavior was only observed in the region where the interfacial tension was finite, and the crossover from $q^2$ to $q^4$ was observed when the interfacial tension was almost zero. In the bilayer membrane, in contrast, the crossover from $q^2$ to $q^4$ in the region where the interfacial tension was finite, while $q^4$ behavior was only observed when the interfacial tension was almost zero.}
		\begin{tabular}[]{cccc} \hline
			          & $q^2$ only                                                            & Crossover $q^2$ to $q^4$                                              & $q^4$ only                                                          \\ \hline
			Monolayer & \raisebox{-0.5\height}{\includegraphics[scale=0.65]{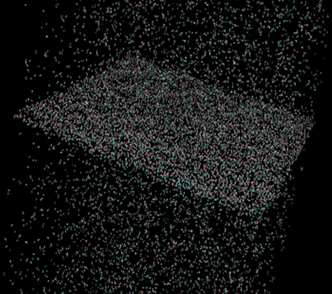}} & \raisebox{-0.5\height}{\includegraphics[scale=0.65]{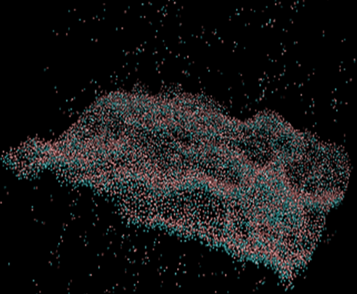}} &                                                                     \\
			          & $\gamma\neq0$                                                         & $\gamma=0$                                                            &                                                                     \\
			Bilayer   &                                                                       & \raisebox{-0.5\height}{\includegraphics[scale=0.15]{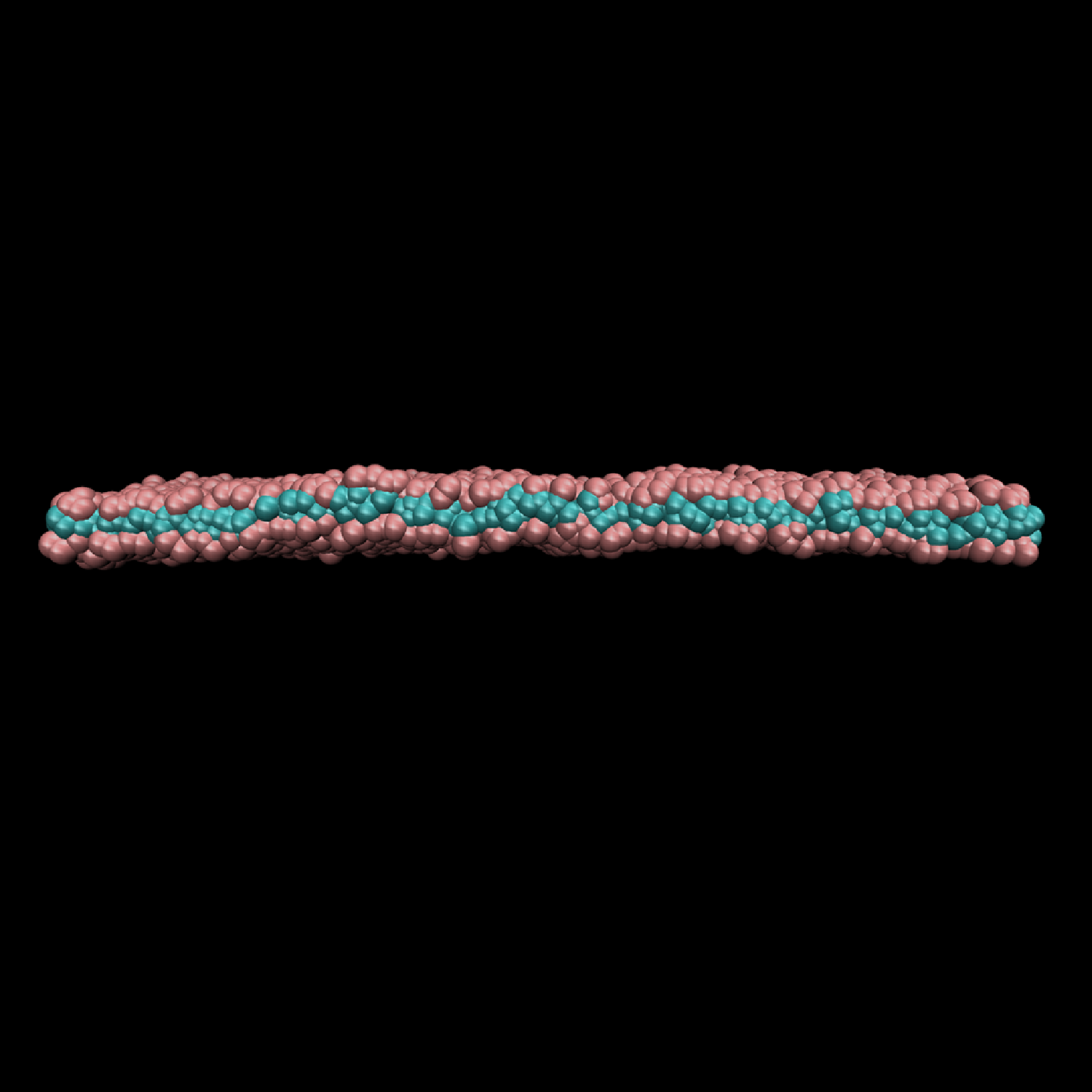}}   & \raisebox{-0.5\height}{\includegraphics[scale=0.15]{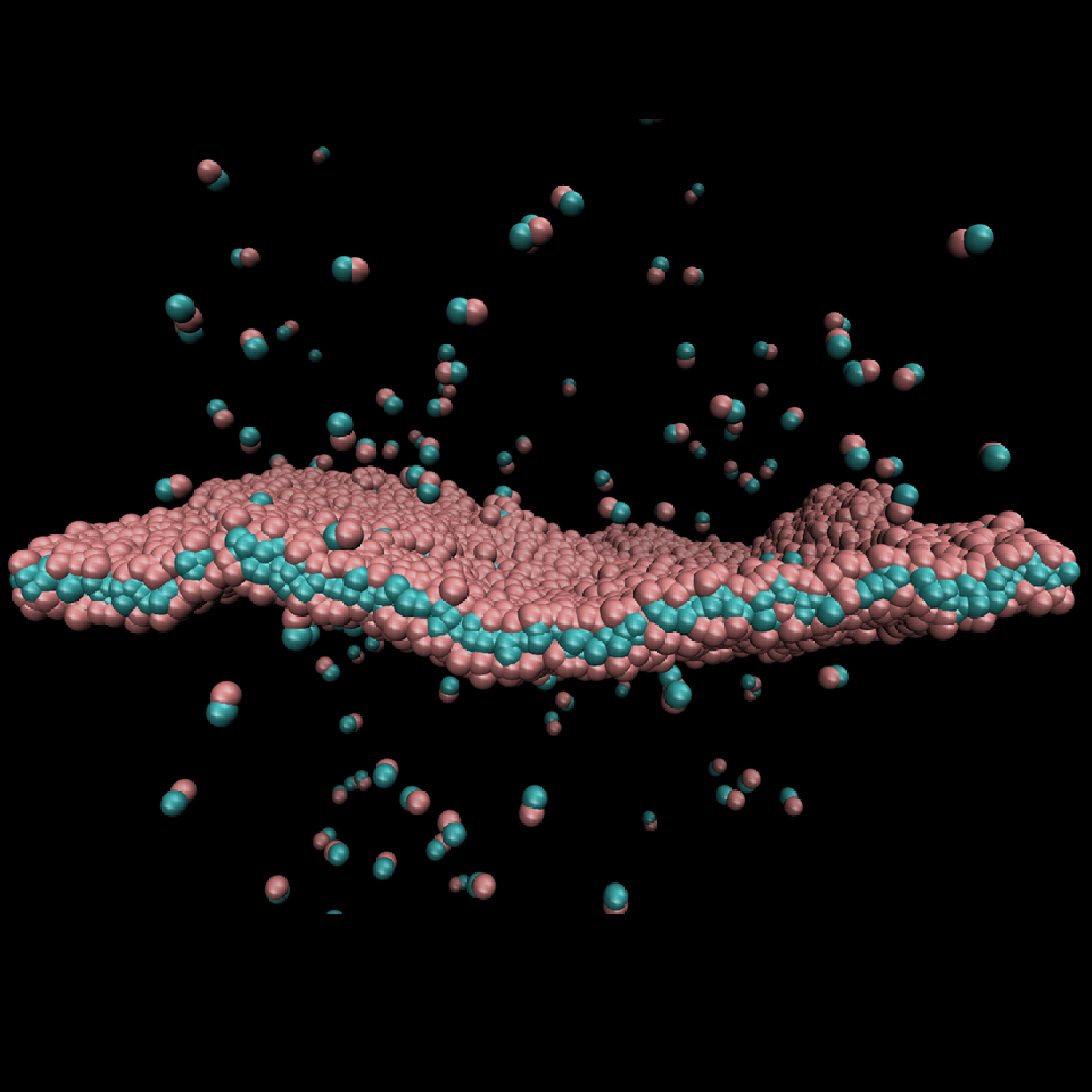}} \\
			          &                                                                       & $\gamma\neq0$                                                         & $\gamma=0$                                                          \\\hline
			\label{table:crossover}
		\end{tabular}
	\end{center}
\end{table}

We also observed RDFs to investigate the relationship between membrane fluctuation and microscopic atomic interactions. We found that the interaction between the same types of atoms depends on the temperature. In contrast, the interaction between different amphipathic molecules did not depend on temperature. These results imply that the bilayers are anisotropic in their interactions, and the anisotropy alters the structure of the membrane. The effect of the asymmetry on the fluctuations is the subject of future work.

Although we modeled bilayers as symmetric membranes, the \textit{real} lipid bilayers are composed of various lipids, and asymmetry facilitates deformation in specific directions\cite{op1979lipid}. In bilayers consisting of diatomic molecules, the effect of the interfacial tension was observed, whereas, in real phospholipids, the resilience due to elasticity is dominant. This difference may be attributed to the long hydrophobic groups of the phospholipids. The future issue is to investigate how the effect of interfacial tension changes when the hydrophobic groups are lengthened. Understanding their behaviors will lead to further insights into lipid bilayers.

\begin{acknowledgment}
	The authors would like to thank H. Noguchi and H. Nakano for fruitful discussions. This research was supported by JSPS KAKENHI, Grant No. JP21K11923. The computation was partly carried out using the facilities of the Supercomputer Center, Institute for Solid State Physics (ISSP), University of Tokyo.
\end{acknowledgment}

\bibliographystyle{jpsj}
\bibliography{reference}

\begin{thebibliography}{10}

\bibitem{nagle2000structure}
J.~F. Nagle and S.~Tristram-Nagle: Biochimica et Biophysica Acta (BBA)-Reviews
  on Biomembranes {\bfseries 1469} (2000) 159.

\bibitem{bloom1991physical}
M.~Bloom, E.~Evans, and O.~G. Mouritsen: Quarterly Reviews of Biophysics
  {\bfseries 24} (1991) 293.

\bibitem{peetla2009biophysical}
C.~Peetla, A.~Stine, and V.~Labhasetwar: Molecular Pharmaceutics {\bfseries 6}
  (2009) 1264.

\bibitem{goetz1998computer}
R.~Goetz and R.~Lipowsky: The Journal of Chemical Physics {\bfseries 108}
  (1998) 7397.

\bibitem{goetz1999mobility}
R.~Goetz, G.~Gompper, and R.~Lipowsky: Physical Review Letters {\bfseries 82}
  (1999) 221.

\bibitem{marrink2003molecular}
S.~J. Marrink and A.~E. Mark: Journal of the American Chemical Society
  {\bfseries 125} (2003) 15233.

\bibitem{Shinoda2010}
W.~Shinoda, R.~DeVane, and M.~L. Klein: The Journal of Physical Chemistry B
  {\bfseries 114} (2010) 6836.

\bibitem{coppock2009atomistic}
P.~S. Coppock and J.~T. Kindt: Langmuir {\bfseries 25} (2009) 352.

\bibitem{venable1993molecular}
R.~M. Venable, Y.~Zhang, B.~J. Hardy, and R.~W. Pastor: Science {\bfseries 262}
  (1993) 223.

\bibitem{orsi2010permeability}
M.~Orsi and J.~W. Essex: Soft Matter {\bfseries 6} (2010) 3797.

\bibitem{fabrikant2009computational}
G.~Fabrikant, S.~Lata, J.~D. Riches, J.~A. Briggs, W.~Weissenhorn, and M.~M.
  Kozlov: PLoS Computational Biology {\bfseries 5} (2009) e1000575.

\bibitem{viallat2014red}
A.~Viallat and M.~Abkarian: International Journal of Laboratory Hematology
  {\bfseries 36} (2014) 237.

\bibitem{yu2017review}
H.~Yu, S.~Engel, G.~Janiga, and D.~Th{\'e}venin: Artificial Organs {\bfseries
  41} (2017) 603.

\bibitem{de1997deformation}
K.~De~Haas, C.~Blom, D.~Van~den Ende, M.~H. Duits, and J.~Mellema: Physical
  Review E {\bfseries 56} (1997) 7132.

\bibitem{zgorski2019surface}
A.~Zgorski, R.~W. Pastor, and E.~Lyman: Journal of Chemical Theory and
  Computation {\bfseries 15} (2019) 6471.

\bibitem{enkavi2019multiscale}
G.~Enkavi, M.~Javanainen, W.~Kulig, T.~R{\'o}g, and I.~Vattulainen: Chemical
  reviews {\bfseries 119} (2019) 5607.

\bibitem{leonard2019developing}
A.~N. Leonard, E.~Wang, V.~Monje-Galvan, and J.~B. Klauda: Chemical reviews
  {\bfseries 119} (2019) 6227.

\bibitem{helfrich1973elastic}
W.~Helfrich: Z. Naturforsch. C {\bfseries 28} (1973) 693.

\bibitem{noguchi2009membrane}
H.~Noguchi: Journal of the Physical Society of Japan {\bfseries 78} (2009)
  041007.

\bibitem{venturoli2006mesoscopic}
M.~Venturoli, M.~M. Sperotto, M.~Kranenburg, and B.~Smit: Physics reports
  {\bfseries 437} (2006) 1.

\bibitem{marrink2019computational}
S.~J. Marrink, V.~Corradi, P.~C. Souza, H.~I. Ingolfsson, D.~P. Tieleman, and
  M.~S. Sansom: Chemical reviews {\bfseries 119} (2019) 6184.

\bibitem{cooke2005tunable}
I.~R. Cooke, K.~Kremer, and M.~Deserno: Physical Review E―Statistical,
  Nonlinear, and Soft Matter Physics {\bfseries 72} (2005) 011506.

\bibitem{brannigan2005flexible}
G.~Brannigan, P.~F. Philips, and F.~L. Brown: Physical Review E―Statistical,
  Nonlinear, and Soft Matter Physics {\bfseries 72} (2005) 011915.

\bibitem{Brandt2011}
E.~G. Brandt, A.~R. Braun, J.~N. Sachs, J.~F. Nagle, and O.~Edholm: Biophys. J
  {\bfseries 100} (2011) 2104.

\bibitem{Kikuchi2023}
S.~Kikuchi and H.~Watanabe: The Journal of Chemical Physics {\bfseries 158}
  (2023).

\bibitem{humphrey1996vmd}
W.~Humphrey, A.~Dalke, and K.~Schulten: Journal of Molecular Graphics
  {\bfseries 14} (1996) 33.

\bibitem{LAMMPS1995}
S.~Plimpton: Journal of Computational Physics {\bfseries 117} (1995) 1.

\bibitem{nose1984molecular}
S.~Nos{\'e}: Molecular Physics {\bfseries 52} (1984) 255.

\bibitem{hoover1985canonical}
W.~G. Hoover: Physical Review A {\bfseries 31} (1985) 1695.

\bibitem{andersen1980molecular}
H.~C. Andersen: The Journal of Chemical Physics {\bfseries 72} (1980) 2384.

\bibitem{verlet1967computer}
L.~Verlet: Physical Review {\bfseries 159} (1967) 98.

\bibitem{watson2012determining}
M.~C. Watson, E.~G. Brandt, P.~M. Welch, and F.~L. Brown: Physical Review
  Letters {\bfseries 109} (2012) 028102.

\bibitem{op1979lipid}
J.~A. Op~den Kamp: Annual Review of Biochemistry {\bfseries 48} (1979) 47.

\end{thebibliography}

\end{document}